\documentclass[twoside]{article}

\usepackage{PRIMEarxiv}

\usepackage[utf8]{inputenc} 
\usepackage[T1]{fontenc}    
\usepackage{hyperref}       
\usepackage{url}            
\usepackage{booktabs}       
\usepackage{amsfonts}       
\usepackage{nicefrac}       
\usepackage{microtype}      
\usepackage{lipsum}
\usepackage{fancyhdr}       
\usepackage{graphicx}       
\graphicspath{{media/}}     

\usepackage{natbib}

\title{How to out-perform default random forest regression: choosing hyperparameters for applications in large-sample hydrology}

\author{
  Divya K. Bilolikar \\
  Department of Mathematics\\
  University of British Columbia \\
  Vancouver, Canada\\
  \texttt{dbilolikar@gmail.com} \\
   \And
  Aishwarya More \\
  Department of Statistics \\
  University of British Columbia \\
  Vancouver, Canada\\
  \texttt{aish.r.more@gmail.com} \\
  \And
  Aella Gong \\
  Independent Researcher\\
  Vancouver, Canada\\
  \texttt{aellaayx@gmail.com}
  `\And
  Joseph Janssen \\
  Department of Earth, Ocean, and Atmospheric Sciences \\
  University of British Columbia \\
  Vancouver, Canada\\
  \texttt{joejanssen@eoas.ubc.ca} \\
}

\begin{document}
\maketitle

\begin{abstract}
Predictions are a central part of water resources research. Historically, physically-based models have been preferred; however, they have largely failed at modeling hydrological processes at a catchment scale and there are some important prediction problems that cannot be modeled physically. As such, machine learning (ML) models have been seen as a valid alternative in recent years. In spite of their availability, well-optimized state-of-the-art ML strategies are not being widely used in water resources research. This is because using state-of-the-art ML models and optimizing hyperparameters requires expert mathematical and statistical knowledge. Further, some analyses require many model trainings, so sometimes even expert statisticians cannot properly optimize hyperparameters. To leverage data and use it effectively to drive scientific advances in the field, it is essential to make ML models accessible to subject matter experts by improving automated machine learning resources. ML models such as XGBoost have been recently shown to outperform random forest (RF) models which are traditionally used in water resources research. In this study, based on over 150 water-related datasets, we extensively compare XGBoost and RF. This study provides water scientists with access to quick user-friendly RF and XGBoost model optimization.
\end{abstract}


\section{Introduction}
Predictions are vital in water resources research. Hydrologists must predict a variety of hydrological quantities such as (1) how much water is available to determine groundwater recharge and water allocation rights, (2) when will water be available for the planning of restoration and hydropower production, (3) how severe will droughts be to assess ecological health and drought risk, and (4) how severe will floods be so that engineers can design better dams, levees, culverts, and reservoirs \citep{bloschl2013runoff}. Since most catchments are ungauged, predicting the previously listed hydrological responses is especially important in ungauged basins. In fact, prediction in ungauged basins was a decade-long prioritized initiative set by the International Association of Hydrological Sciences in 2003.  

Physically-based models and statistical models have been used to make hydrological predictions for both gauged and ungauged basins. While physically-based models have dominated hydrological modeling historically, they have many downsides \citep{beven1989changing,lange2020machine}. For example, some physical processes such as Richards equation are inappropriately applied for large grid cells, when it is only applicable to sub-meter scales \citep{beven1989changing,lange2020machine}. Computational resources such as time and memory constrain our ability to reduce grid sizes, but even if these issues are solved, equifinality, over-parameterization, and too little data can produce poor models with little extrapolative power \citep{bloschl2013runoff,halford2004more,lange2020machine,beven1989changing}. Moreover, important problems such as precipitation infilling \citep{tang2020scdna}, water use prediction \citep{chang2017determinants}, and snow depth to snow water equivalent conversion \citep{ntokas2021investigating}, cannot be modeled physically. 

Since hydrological problems are complex, better modelling strategies are essential. Model selection is a formidable challenge for water managers and scientists. For each problem they encounter, the trade-offs between fidelity, complexity, and resources (e.g., time, skill, data, and money) must be considered \citep{golden2017integrating}. Even when limiting the selection to only statistical models, there are many to choose from. Models such as multiple linear regression, lasso, and elastic net are simple and easy to compute, but give poor predictions for complex, highly nonlinear hydrological problems. Modeling approaches which use machine learning (ML) such as support vector machines, neural networks, tree-based methods, and boosting, which can learn high-order, nonlinear interactions, have also been used in hydrology. Random forests (RF), a tree-based method, is one of the most popular machine learning models in hydrology \citep{tyralis2019brief,lange2020machine}. 

Over twenty years have passed since Breiman’s seminal paper introducing RF was published \citep{breiman2001random}. With ML being such a new and quickly expanding area of research, one would think that statisticians and computer scientists would have developed more advanced ML methods for better predictions since 2001, and scientists in other fields would be using these more advanced methods instead of RF. While the former is true, the latter is not, at least in hydrology. In fact, while RF models were not extensively used in water resources research until almost 15 years after their conception, the use of RF in water resources research is now increasing exponentially \citep{tyralis2019brief}. RF is an extremely powerful method that can model complex nonlinear relationships without overfitting, making it easy to use and applicable to a wide range of applications \citep{breiman2001random}. Although RF performs better at classification when compared to regression, water researchers mainly use RF for regression applications \citep{breiman2001random,tyralis2019brief}. Therefore, in this paper, we will focus on regression.  

When training RF models, users have the ability to change the structure and randomness of the model by tuning hyperparameters. These hyperparameters, when optimized, can increase the predictive strength of the model \citep{probst2019tunability}. Though hyperparameter optimization (HPO) is recommended, hydrologists often skip this step or test a small set of hyperparameters due to limited time resources. For example, \citet{tang2020scdna} used RF to impute missing precipitation and temperature data across North America to great effect. In fact, even though the hyperparameters were set arbitrarily, RF outperformed neural networks. Likewise, \citet{addor2018ranking} did not perform HPO and used default hyperparameters to predict hydrological signatures based on catchment attributes. \citet{teweldebrhan2020coupled} used RF to identify parameters for a distributed hydrological model. In this study, only one of RF’s five hyperparameters was optimized. Further, RF has been trained to predict short-term daily streamflow by \citet{pham2021evaluation}. Again, only one hyperparameter was tuned. Groundwater withdrawal estimates were successfully made over the central United States using evapotranspiration, precipitation, land use, and crop coefficient with RF by optimizing two hyperparameters \cite{majumdar2020groundwater}. 

Extreme gradient boosting (XGBoost) is a recent advancement in machine learning. It is an efficient gradient tree boosting algorithm that iteratively improves predictions by training on residuals. XGBoost is state-of-the-art. In fact, it won 17 of the 29 Kaggle data science competitions in 2015, compared to neural networks which only won 11 \citep{chen2016xgboost}. While HPO has a relatively small impact on the predictive performance of RF models, the impact on XGBoost models is much greater \citep{probst2019tunability}. Performing HPO for XGBoost is also more complex because of the higher dimensional parameter space. Not only can practitioners control the randomness and structure of the model, but also the learning rate and regularization factors can be chosen. XGBoost is not as prevalent in hydrology compared to RF, though there have been some instances of use very recently. XGBoost, for example, has been used for streamflow forecasting \citep{zhang2019dynamic,ni2020streamflow,gauch2021proper,gauch2019data}. With climate variables such as pressure, radiation, temperature, humidity, and wind speed, \citet{bacsaugaouglu2020hybridized} and \citet{fan2018evaluation} used XGBoost to predict daily evapotranspiration values. XGBoost was only outperformed by neural networks but was the most computationally efficient method \citep{fan2018evaluation}. 

Model selection, including choosing the appropriate algorithm and hyperparameters is tedious and requires sufficient knowledge of algorithms, statistics, and machine learning, so this problem is not well suited for hydrologists \citep{zoller2021benchmark}. From a survey of natural resource managers, \citet{saia2020transitioning} found machine learning is not widely used in natural resources management because it seems confusing and risky. Further, no clear description on how to use machine learning toolkits are available for non-experts. In particular, hydrologists assume HPO is extremely computationally costly and has little effect on prediction results \citep{papacharalampous2018univariate,tyralis2021super}. While most applications of machine learning only require a single model, others, such as feature importance, require users to train a huge number of models with slightly different input datasets \citep{janssen2022ultra,covert2020understanding,catav2021marginal,li2022statistical}. This prevents even experts from properly tuning hyperparameters due to the computational burden \citep{janssen2022ultra,catav2021marginal}. While machine learning algorithms are often compared in hydrologically related papers \citep{gauch2021proper,fan2018evaluation,tang2020scdna,worland2018improving}, the number of problems in which they are compared are often limited, so no generalizable conclusions can be made about the superiority of one algorithm over another. Thus, just deciding which machine learning algorithm to use can be costly.

In this study, we take a step towards removing the above barriers by (1) comparing RF with XGBoost on a large set of large-sample hydrology regression problems such that hydrologists can make informed decisions about which algorithm to choose and (2) providing resources for quick and user-friendly RF and XGBoost HPO. To fulfill these goals, we will compare the performance of RF and XGBoost on over 150 hydrologically related regression datasets using performance metrics that are familiar to hydrologists such as Kling-Gupta-Efficiency (KGE) and Nash–Sutcliffe-Efficiency (NSE). Then, we will train hyperparameter optimization meta-learning models such that non-experts can utilize fully optimized RF and XGBoost models with ease.

\section{Data}
\label{sec:Data}

The data used in this paper comes from large-sample hydrology. Large-sample hydrology (LSH) is a branch of hydrology and more specifically comparative hydrology, where all analysis is done using a large selection of catchments (ranging anywhere from tens to thousands of catchments) \citep{addor2020large}. LSH provides multivariate data that can be used in an attempt to generate hydrological knowledge which is generalizable across locations, climates, and scales such that models do not need direct hydrologic information to calibrate models for each stream location \citep{gupta2014large,hrachowitz2013decade}. One advantage of LSH is that with larger samples, there is a greater chance of having hydrology-wide advances. Additionally, having large datasets reduces the error and increases the confidence in being able to make generalizable conclusions since unusual cases may be present \citep{gupta2014large,andreassian2010court}. Extensive testing is essential to quantify the validity of hydrological models. These tests are done on a large sample of unique catchments over a long period of time. If a model is not sufficiently tested then any results from that model cannot be used with a high degree of certainty \citep{linsley1982rainfall}. However, there are some limitations to LSH, namely, there are issues with accessing large datasets. For example, data from a certain country may not be accessible to foreigners, though some globally available datasets do exist \citep{gupta2014large}.

The LSH datasets used in this study had to be cleaned before use. First, rows with missing response variable data are removed. Then, the percentage of missing data in each explanatory column was calculated. If some columns had more than 10\% missing data then two datasets were created, one with all columns with less than 50\% missing data and one with all columns with less than 10\% missing data. This was done to create a more diverse set of data for training the meta learning models. Each dataset then was separated into multiple tables where the first column was a different response variable with the explanatory variables following after it. The number of observations range from 94 to 1366 and the number of explanatory variables range from 3 to 100 (Table \ref{tab:datasets}). During data processing, the zero flow data from Chile, Central Europe, and Great Britain were removed because there were too many zeros and regression was deemed inappropriate. Additionally, CAMELS\_GB high flow duration data was deleted because there were outliers that highly altered the results. All datasets were processed in RStudio version 1.3.1093. 

\begin{table}[h!]
 \caption{Names of the large sample datasets of interest along with the number of observations, the number of explanatory variables, the number of datasets used from each file, the region from which the measurements were taken, and the source.}
 \label{tab:datasets}
  \centering
  \begin{tabular}{llllll}             \\
    \cmidrule(r){1-6}
    Name & \# of Obs & \# of Features & \# of Datasets & Region & Reference \\
    \midrule
    CAMELS\_AUS & 94  & 92 & 13 & Australia & \citep{fowler2021camels}     \\
    CAMELS\_BR & 854  & 50 & 13 & Brazil & \citep{chagas2020camels}     \\
    CAMELS\_CL & 153  & 69 & 16 & Chile &  \citep{alvarez2018camels}      \\
    CAMELS\_GB & 356-669  & 91-100 & 24 & Great Britain & \citep{coxon2020camels}      \\
    CAMELS\_US & 507-643  & 44-45 & 26 & United States & \citep{addor2017camels}     \\
    LSH\_NA & 571  & 3-10 & 30 & North America & \citep{janssen2021hydrologic}     \\
    LamaH & 796  & 61 & 13 & Central Europe & \citep{klingler2021lamah}     \\
    Kuentz\_Europe & 1366  & 49 & 16 & Europe & \citep{kuentz2017understanding}     \\
    \bottomrule
  \end{tabular}
  \label{tab:dataTable}
\end{table}

\section{Methods}
\subsection{Random Forests}
The random forest (RF) machine learning algorithm is an ensemble tree-based method used for modelling complex non-linear relationships without overfitting \citep{breiman2001random}. RF predictions are obtained by aggregating the results from a series of de-correlated and randomized deep decision trees generated via a process known as bootstrap aggregating or \textit{bagging} \citep{svetnik2003random}. Bagging is a model aggregating algorithm where models are trained on random bootstrap samples from the training data, and the aggregated result is used as the final prediction value. Additionally, what makes RF unique from other ensemble methods is that each decision tree uses a random subset of the features to split each node. This is how de-correlated trees are obtained, since they all use a specific and unique node splitting and training data subset \citep{hastie2009random}. 

RF regression models are developed using regression decision trees, while classification decision trees are used to develop random forest classification models. Since hydrologists more commonly use RF regression, we will focus on the implementation of RF for regression accordingly. There are many different implementations of RF across different programming languages. We will be making use of the \emph{ranger} R package to implement the random forest algorithm for our work \citep{wright2015ranger}. 

As with many machine learning algorithms, users have the option to adjust the RF model to adhere to certain characteristics by tuning hyperparameters. The hyperparameters for RF of interest to us include mtry, num.trees, replace, min.node.size, and sample.fraction which are described further in Table \ref{tab:RFHP}. 

\subsection{XGBoost}
Gradient tree boosting is a method that involves training multiple decision trees in a sequential manner such that each additional tree takes into account the loss or error from the previous tree (Friedman, 2001). By implementing gradient boosting, the algorithm uses a series of weak learners to form a single strong learner, thus improving predictions. 

Extreme gradient boosting, or XGBoost, is an ensemble machine learning method that employs gradient tree boosting to produce highly scalable predictive models \citep{chen2016xgboost}. It can be used for both developing classification and regression models, but we will focus on its use for regression problems as that aligns with the scope of our research. Some of the main features of XGBoost that differentiate it from other ensemble tree methods are that it utilizes regularized boosting, runs parallel tree learning, and can accommodate for sparse or missing data. The XGBoost algorithm is also adept at handling sparse data while avoiding overfitting of the data. This is a result of the algorithm using both L1 and L2 regularization based boosting and tree pruning \citep{chen2016xgboost}. Another notable feature of the algorithm is that it implements parallelized sequential tree building. This makes XGBoost highly scalable and good for larger datasets, since it is algorithmically efficient when compared to other tree based ensemble methods. 

For our work, we will be making use of the \emph{xgboost} R package to implement XGBoost \citep{chen2016xgboost}. The algorithm makes use of hyperparameters to define the specifications of the regression model. The hyperparameters that are of interest to us include the step size shrinkage, the subsample ratio of training instances, the maximum depth of trees, the minimum child node weights, the subsample ratio of columns used for each tree, and three regularization parameters including alpha, and lambda (Table \ref{tab:XGBHP}). 

\subsection{Model Evaluation}

\subsubsection{Time Complexity}

The \emph{randomForest} package is the most commonly used R package in hydrology to train RF models \cite{zhang2021global,tyralis2019brief}, while others use ranger \cite{sekulic2021high}. We will compare the training speeds for each machine learning algorithm (package) including XGBoost (\emph{xgboost}) and RF (\emph{randomForest} and \emph{ranger}) \citep{wright2015ranger,liaw2002classification,chen2015xgboost}. We will analyze how training times grow with the number of trees and the number of training observations using subsamples of our Kuentz\_Europe dataset \citep{kuentz2017understanding}. We ran the algorithms on a set of 50 to 5000 trees and increased the number of trees in each trial by 100 every time and used all observations from our dataset for all trials. These trials were then run 10 times each and the runtime of these trials were recorded and the average across replications was computed. We ran the packages on a set of 50 to 1366 samples with increments of 100 while keeping the number of trees constant at 500. Similarly to the first part, each trial was run 10 times and the average of the trails is taken as the final result. These experiments were run on a computer with a 1.4 GHz Quad-Core Intel Core i5 and 16 GB 2133 MHz RAM. 

\subsubsection{Predictive Power}

To compare the effects of using different HPO methods on the RF and XGBoost models trained on the same hydrological data, we identified the need to employ hydrology-specific evaluation methods and metrics in order to make our results as suitable for hydrologists as possible. Thus, we will primarily be using the Nash-Sutcliffe efficiency criterion \citep{nash1970river} and the Kling-Gupta Efficiency criterion \citep{gupta2009decomposition}, along with 10-fold cross validation to assess and evaluate model performance. Cross validation is a technique to evaluate predictive models by partitioning the original sample into a training set to train the model, and a test set to evaluate it \citep{browne2000cross}. We will use 20\% of the data as our test set exclusively, while the remaining data is used as our training set for cross validation and choosing hyperparameters. Using 10-fold cross validation here means the data is split into ten equally sized subsets, where the calculated KGE and NSE act as the evaluation metric for each split. 

The Nash-Sutcliffe efficiency is defined as:

\begin{equation}
NSE= 1 - \frac{\sum{(y-\hat{y})^2}}{\sum{(y-\bar{y})^2}}.
\end{equation}

The NSE value is calculated as one minus the ratio of the error variance of the modeled values to the variance of the observed values. Here, $\hat{y}$ refers to the predicted values from the model, $y$ refers to the observed values, and $\bar{y}$ is the mean of the observed values.

A model with an NSE score equal to one is considered to be the perfect model, while a model with an NSE score equal to zero is shown to have estimation error variance equal to the variance of the observed model \citep{mccuen2006evaluation}. If the NSE is negative, then the mean of observed values is a better predictor than the assessed model. So, we would reject any model with an NSE score that is not between zero and one. 

The modified Kling-Gupta Efficiency is defined as: 

\begin{equation}
KGE= 1 - \sqrt{(r-1)^2 + (\alpha -1 )^2 + (\beta -1)^2}.
\end{equation}

The KGE acts as a measure of the trade-off between the correlation, bias, and variability components represented in the hydrological model. The first component, $r$, is the correlation between the observations and predictions, the second component, $\alpha$, is the ratio between the mean predictions versus the mean of the observations, and the third component, $\beta$, is the ratio between the coefficient of variation of the predictions and the coefficient of variation of the observations. These three components define a three-dimensional criteria space, then the KGE objective is defined as one minus the distance from the optimal point to the Pareto front. Hydrologists generally use the Kling-Gupta Efficiency as an evaluation metric since it improves upon the shortcomings of NSE including underestimating variability and scaled biases \citep{gupta2009decomposition}. The modified Kling-Gupta Efficiency further improves upon the original KGE measure by de-correlating the bias and variability components \citep{kling2012runoff}. The KGE value can range between negative infinity and one, where we say the model is more accurate the closer the value gets to one.

\subsection{Hyperparameter Optimization Algorithms}

The structure and effectiveness of a machine learning model are determined by how its hyperparameters are set. To obtain the best configuration of hyperparameter values, many resort to automated HPO as it reduces human effort, reduces any possible bias, and improves the ML algorithm’s performance \citep{feurer2019hyperparameter}. While HPO tuning strategies could be implemented using traditional optimization techniques such as gradient descent, this would not be efficient as it is not common for HPO problems to be convex or differentiable in nature \citep{yang2020hyperparameter}. HPO algorithms such as grid search and random search are a few of the common and popular HPO algorithms, with which some hydrologists would be familiar \citep{ni2020streamflow,gauch2021proper,feigl2021machine,zhang2019dynamic}. Below we describe the different HPO methods we compare in our paper including default hyperparameters, optimal default hyperparameters, random search, and our newly proposed hyperparameter meta-learning algorithm. 

\begin{table}[h!]
 \caption{Names of random forest hyperparameters of interest to us along with their data type (Type), their respective upper and lower bounds (Range), their default setting (Default), their optimal default setting (Opt Default), and their random search space (Random Search).}
 \label{tab:RFHP}
  \centering
  \begin{tabular}{llllll}             \\
    \cmidrule(r){1-6}
    Name & Type & Range & Default & Opt Default & Random Search \\
    \midrule
    mtry & Numeric  & $(0-1)*p$ & $\sqrt{p}$ & $0.257p$ & $[0,1]*p$     \\
    num.trees & Integer  & $1-\infty$ & $500$  & $983$ & $[10,2000]$     \\
    replace & Boolean  & $\{0,1\}$ & 1 & 0 &  $\{0,1\}$      \\
    min.node.size & Numeric  & $1-n$ & 0 & 1 & $n^{[0,1]}$      \\
    sample.fraction & Numeric  & $0-1$ & 1 & 0.703 & $[0.1,1]$     \\
    \bottomrule
  \end{tabular}
\end{table}

\begin{table}[h!]
 \caption{Names of xgboost hyperparameters of interest to us along with their data type (Type), their respective upper and lower bounds (Range), their default setting (Default), their optimal default setting (Opt Default), and their random search space (Random Search).}
 \label{tab:XGBHP}
  \centering
  \begin{tabular}{llllll}             \\
    \cmidrule(r){1-6}
    Name & Type & Range & Default & Opt Default & Random Search \\
    \midrule
    nrounds & Integer  & $[1,\infty)$ & $500$ & $4168$ & $[1,5000]$ \\
    eta & Numeric  & $(0,1)$ & $0.3$ & $0.018$ & $2^{[-10,0]}$     \\
    subsample & Numeric  & $(0,1]$ & $1$ & $0.839$ & $[0.1,1]$    \\
    max\_depth & Integer & $[1,\infty)$ & $6$ & $13$ &  $[1,15]$      \\
    min\_child\_weight & Numeric  & $(0,\infty)$ & $1$ & $2.06$ & $2^{[0,7]}$      \\
    colsample\_bytree & Numeric  & $(0,1]$ & $1$ & $0.752$ & $(0,1]$     \\
    alpha & Numeric  & $(0,\infty)$ & $1$ & $1.113$ & $2^{[-10,10]}$     \\
    lambda & Numeric  & $(0,\infty)$ & $1$ & $0.982$ & $2^{[-10,10]}$     \\
    \bottomrule
  \end{tabular}
\end{table}

\subsubsection{Default Hyperparameters}

Default hyperparameters are generally obtained by empirical experiments using numerous, varied datasets \citep{probst2019tunability}. This is done to identify the hyperparameter configuration that will work best with an arbitrary dataset. The main benefit of using default hyperparameters is that they allow for relatively good model performance with no added computational expense. However, the model prediction accuracy may suffer when using default value since they do not account for the problem dimensionality (size of dataset) or the specific features of the dataset \citep{probst2019tunability}. In the ranger R package, the default hyperparameters are as follows: mtry=sqrt(p), num.trees=500, replace=TRUE, min.node.size=0, and sample.fraction=1 (Table \ref{tab:RFHP}). In the xgboost R package, the default hyperparameters are defined as follows: eta=0.3, subsample=1, max\_depth=6, min\_child\_weight=1, colsample\_bytree=1, alpha=0, and lambda=1 (Table \ref{tab:XGBHP}). 

\subsubsection{Optimal Default Hyperparameters}

In a manner similar to ours, previous researchers set out to find optimal default hyperparameter such that they could simplify HPO for less experienced users \citep{probst2019tunability}. They tuned hyperparameters on a set of 38 open-source classification datasets, then found the hyperparameter set that minimizes the average AUC across their datasets \citep{probst2019tunability}. Their new optimal defaults were shown to significantly outperform the package defaults \citep{probst2019tunability}. The optimal default hyperparamters found for the ranger package were as follows: mtry=0.257p, num.trees=983, replace=FALSE, min.node.size=1, and sample.fraction=0.703 (Table \ref{tab:RFHP}). For the xgboost package, the optimal defaults were: eta=0.018, subsample=0.839, max\_depth=13, min\_child\_weight=2.06, colsample\_bytree=0.752, lambda=0.982, and alpha=1.113 (Table \ref{tab:XGBHP}). 

\subsubsection{Random Search}

Random search is a search-based hyperparameter tuning strategy, where hyperparameter configurations are randomly selected from a given search domain. It was proposed as a better alternative to grid search and manual search methods as it is able to find similar results in a fraction of the computation time and is more scalable \citep{bergstra2012random}. Since we are looking to find the best hyperparameter configurations, we will be running our random search with values that fall in the tuning space determined by \citet{probst2019tunability}. For random forests, our search space is mtry $\in [0.1,1]$, num.trees $\in [10,2000]$, replace $\in \{TRUE, FALSE\}$, min.node.size $\in n^{[0,1]}$, and sample.fraction $\in [0.1,1]$ (Table \ref{tab:RFHP}). For xgboost, our search space goes as follows: eta $\in 2^{[-10 ,0]}$, subsample $\in [0.1,1]$, max\_depth $\in [1,15]$, min\_child\_weight $\in 2^{[0,7]}$, colsample\_bytree $\in [0,1]$, lambda $\in 2^{[-10,10]}$, and alpha $\in 2^{[-10,10]}$ (Table \ref{tab:XGBHP}). 

\subsubsection{Meta-model}

Many studies have focused on meta-learning and AutoML with the goal of reducing computational complexity and making ML more suitable for non-experts. The lack of use of HPO by hydrologists can be mitigated by improving automated machine learning (AutoML) resources and providing better optimal defaults via meta-learning \citep{ledell2020h2o,feurer2015efficient}. AutoML automates various parts of the machine learning pipeline which can make ML models easy to build without sacrificing optimality and robustness, allowing hydrologists to focus on the analysis of results \citep{thornton2013auto,ledell2020h2o}. Meta-learning studies can accelerate the HPO task while making the process easier for non-experts by generalizing the relationship between optimal hyperparameters and dataset characteristics from previous learning tasks \citep{feurer2015efficient,vanschoren2019meta}. Our meta-model implements meta-learning techniques, which is when we use machine learning methods to predict the most appropriate parameters or model for a machine learning problem \citep{cui2016recommendation}. In our case, we are using meta-learning to determine the best hyperparameters to use in a RF or XGBoost model based on the dataset we pass to our meta-model.  

\begin{figure}
  \centering
  \includegraphics[width=1\textwidth]{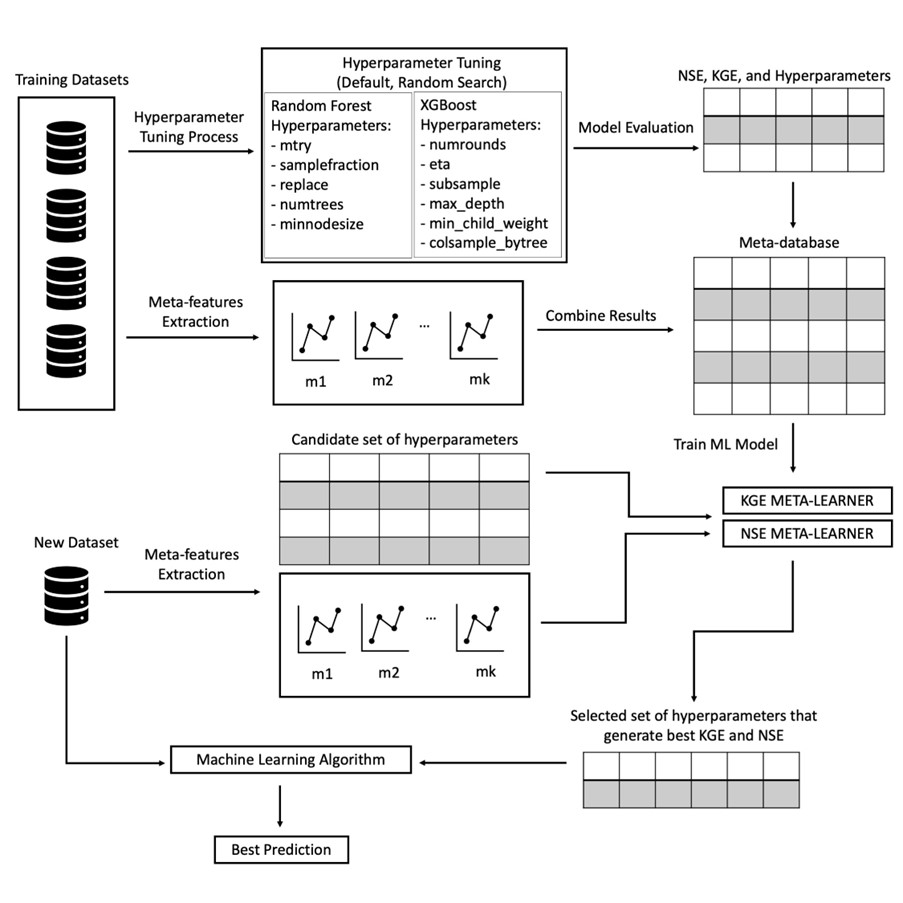}
  \caption{Description of each step taken to train and use our meta-models.}
  \label{fig:meta-model}
\end{figure}

The process for generating the meta-database and using the meta-database to extract hyperparameter predictions is illustrated in Figure \ref{fig:meta-model}. To generate the meta-database, which is used to train the meta-learners, a model is trained and tested using default, optimal default, and 100 iterations of random search hyperparameters for each ML algorithm. The hyperparameters used in each iteration, the meta-features of the training datasets, and the obtained KGE and NSE values from each iteration on the test set, are saved as data in the meta-database. The complexity and general metadata is extracted using the \emph{ECol} \citep{lorena2019complex} and \emph{mfe} \citep{alcobacca2020mfe} R packages respectively. This meta-database is then used to train two models per ML algorithm, one that predicts standardized KGE and another that predicts standardized NSE, based on the hyperparameters and metadata that make up the meta-database. We repeat this process without the metadata such that we can learn the best hyperparameters for a general dataset.

Given a new dataset, the meta-features of the dataset are extracted and passed along with candidate sets of hyperparameters to the meta-learner. This returns the optimal hyperparameter configuration as the one that is predicted to generate the highest standardized KGE or NSE. Finally, this optimal set of hyperparameters is used with the machine learning algorithm to obtain the best predictions for our dataset. 

\section{Results}

\subsection{Time Complexity}

\begin{figure}
  \centering
  \includegraphics[width=0.4965\textwidth]{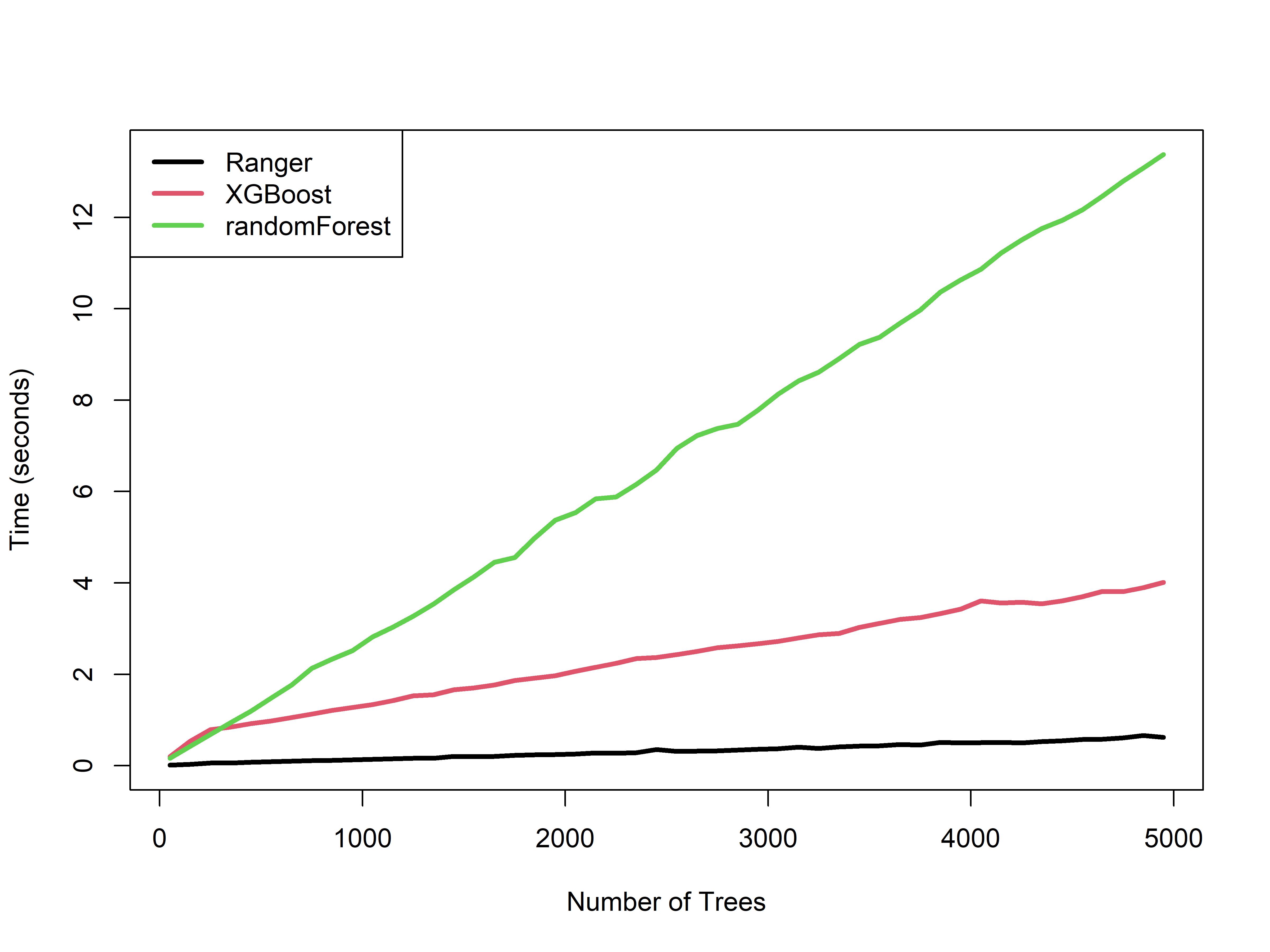}
  \includegraphics[width=0.4965\textwidth]{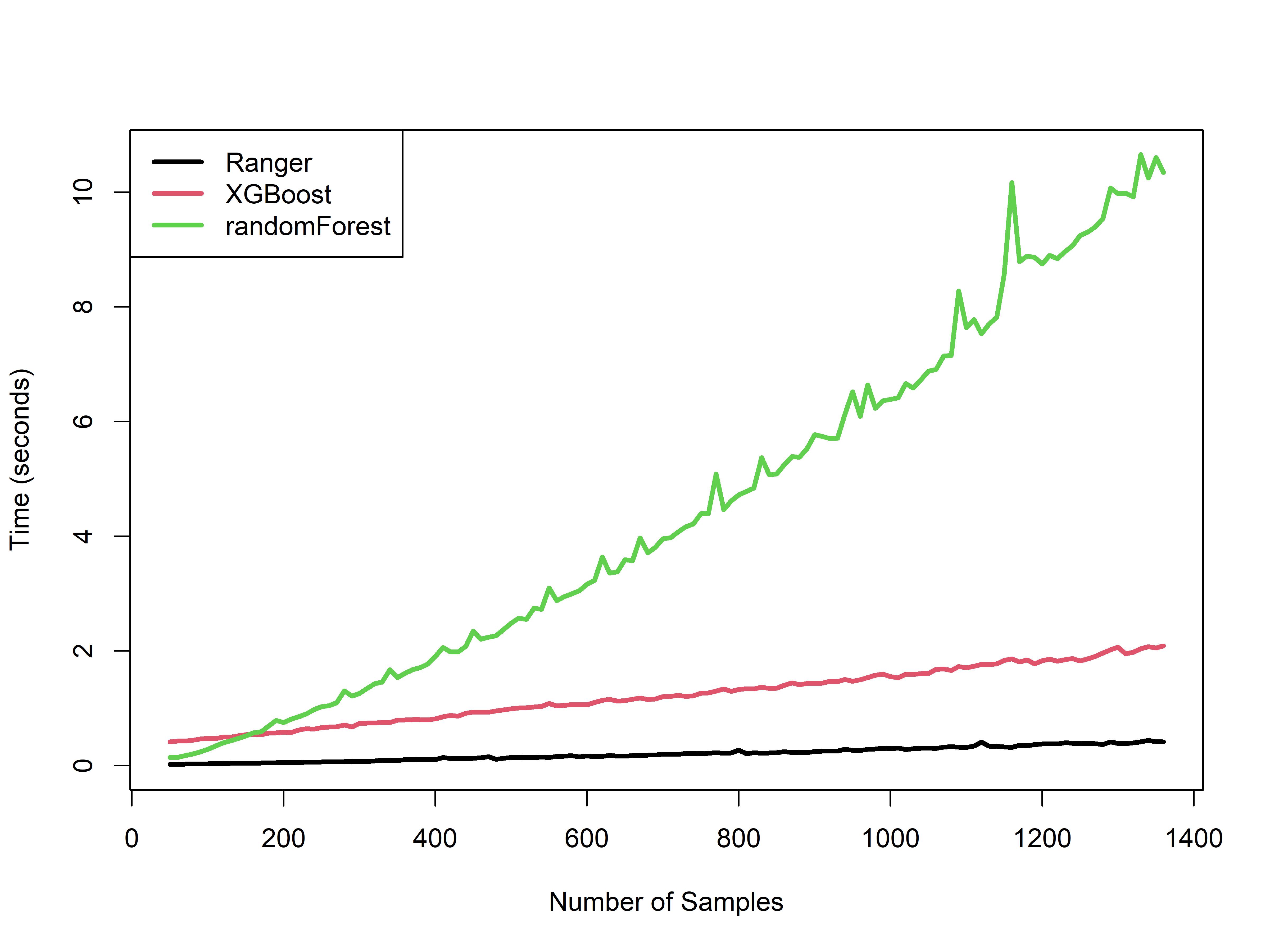}
  \caption{Comparing time complexities of XGBoost (xgboost) and two random forests R packages (randomForest and ranger) as the number of trees or the number of samples increases.}
  \label{fig:time}
\end{figure}

In Figure \ref{fig:time} the time complexities of randomForest , ranger and xgboost are shown as a function of the number of trees used and the number of samples in the training set. In the figure on the left, we can observe that all three packages train models with 50-800 trees with a similar runtime. As the number of trees increases, the runtime for randomForest increases steeply at a linear rate while both ranger and xgboost increase more slowly. We can see that even though randomForest and ranger both implement RF, there is a large discrepancy in their time complexities likely because ranger uses more efficient node splitting techniques \citep{wright2015ranger}. Beyond 1000 trees randomForest  is more than twice as slow as xgboost and more than four times slower than ranger. Although the time complexity for ranger and xgboost is much less than randomForest , there are differences in speed between the two packages. Up to 200 trees, the time complexity of xgboost increases rapidly, but beyond 200 trees the increase is much slower. Indeed, ranger is consistently the fastest algorithm in Figure \ref{fig:time}. 

The plot on the right of Figure \ref{fig:time} shows that ranger is the fastest algorithm for all numbers of samples, while randomForest and xgboost have a similar but slower runtimes, with xgboost being slightly slower when there are less than 150 samples. After 150 samples, randomForest begins to become slower than xgboost. Beyond 1000 samples xgboost is more than three times faster than randomForest. All three algorithms seem to have fairly linear relationships in Figure \ref{fig:time}, though the runtime randomForest may rise superlinearly with the number of samples. 

Figure \ref{fig:time} indicates that the performance of xgboost is better than randomForest  and rapidly becomes better beyond 800-1000 trees or samples. Figure \ref{fig:time} also indicates that ranger is consistently better than xgboost. From both graphs above we can see the importance of optimizing a package, even though randomForest and ranger both use RF they perform very differently.

\subsection{Predictive Power}

\begin{figure}[h!]
  \centering
  \includegraphics[width=0.95\textwidth]{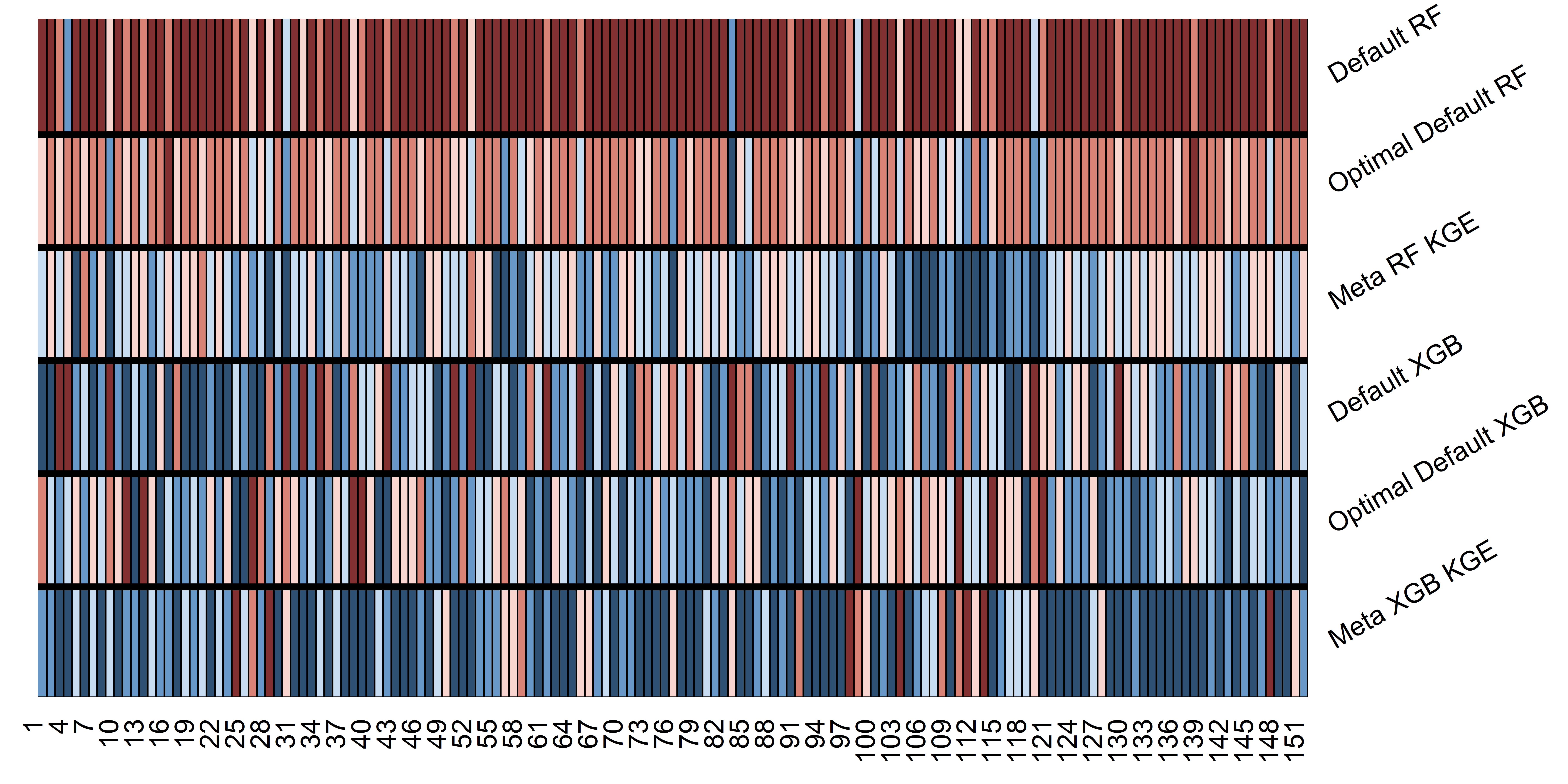}
  \caption{A visual representation of the ranks of each method according to their test set KGE score. Each column represents one of 151 datasets. Each row represents one of six machine learning and hyperparameter optimization methods. The colors represent the ranks of each method for a specific dataset, where dark blue is the method with the highest (best) KGE score and dark red is the method with the lowest (worst) KGE score.}
  \label{fig:KGEresults}
\end{figure}

In order to visualize and compare the best optimizations methods and the predictive power of each optimization method, a heatmap analysis was used. Figure \ref{fig:KGEresults} shows that when KGE is used an the evaluation metric, RF with default hyperparameters has the most dark red rectangles (117) and thus this is the comparatively worst method, especially since it achieved the best results for exactly zero out of the 151 datasets. The optimal default hyperparameters of RF consistently produced the second worst results, with this method only achieving the best result once and the worst result twice. While RF with hyperparameters from our meta-model produce the best results less often compared to all XGB attempts (20 datasets), it is the only method that never produces the worst result indicating that is may be considered as the safe choice in general. The default hyperparameters for XGB provide good but unstable results with this method achieving the best results on 33 datasets and performing the worst on 16 datasets. The optimal default hyperparameters for XGB suffer from the same issue with 24 best results and 9 worst results. It can also be seen that XGBoost with hyperparameters from our meta-model has the most dark blue rectangles (73) and thus is the best optimization method (only 7 dark red rectangles). 

\begin{figure}[h!]
  \centering
  \includegraphics[width=0.95\textwidth]{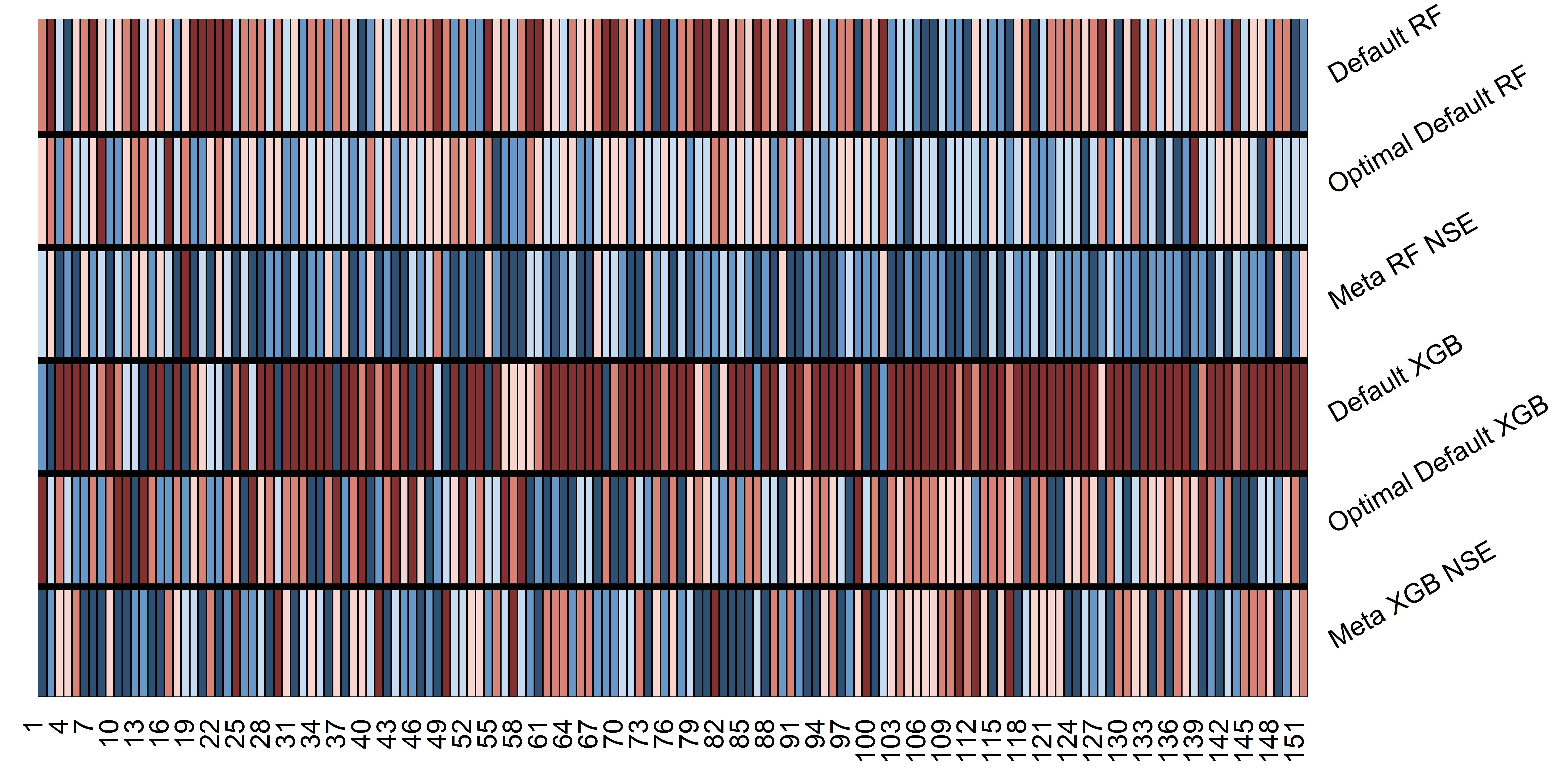}
  \caption{A visual representation of the ranks of each method according to their test set NSE score. Each column represents one of 151 datasets. Each row represents one of six machine learning and hyperparameter optimization methods. The colors represent the ranks of each method for a specific dataset, where dark blue is the method with the highest (best) NSE score and dark red is the method with the lowest (worst) NSE score.}
  \label{fig:NSEresults}
\end{figure}

Figure \ref{fig:NSEresults} shows that XGB with default hyperparameters produces the worst NSE results (98 worst results and 16 best results), while RF with meta-model hyperparameters produced the best NSE results (50 best results and 1 worst result). RF with default hyperparameters also compares poorly with the method achieving the best result only 11 times while achieving the worst result on 25 datasets. RF with optimal default hyperparameters provides consistently middling results since it performs the best on only 7 datasets and the worst on only 3 datasets. The optimal default XGB method is inconsistent (27 best results and 14 worst results), while the meta XGB method is good (40 best results and 10 worst results) but still fails to outperform the meta RF method. In a similar manner to the KGE results we find that RF with hyperparameters predicted from our meta-model is the safe option with the least amount of the lowest ranking results.  

\begin{figure}[h!]
  \centering
  \includegraphics[width=0.99\textwidth]{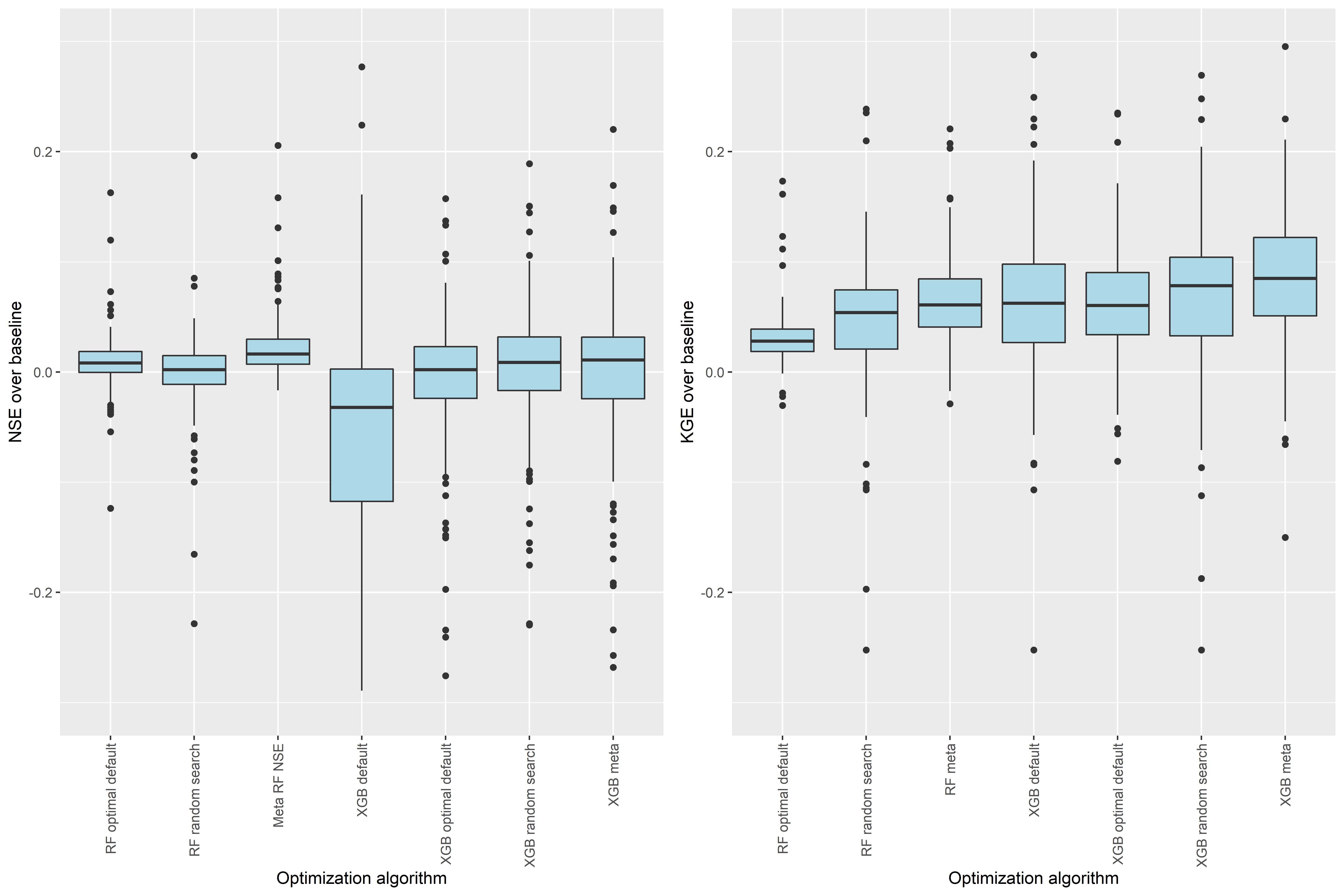}
  \caption{The increase in NSE (left) and KGE (right) that each optimization algorithm can achieve above the baseline method (random forests with default hyperparameters).}
  \label{fig:boxplot}
\end{figure}

Figure \ref{fig:boxplot} shows us how much each method can outperform default random forests, quantified by the increase in KGE and NSE scores. As we can see, all score increases are between negative and positive $0.3$. Focusing on the NSE results, we see that optimal default and meta-model RF consistently outperforms default RF, while default XGB consistently performs worse than default RF. The consistent obtainable increases in NSE are small, though occasionally we observe increases by over 0.1 which is highly significant. To obtain such large increases in NSE, users would likely have to try all methods before finding one that is substantially better than default RF, pointing to the importance of trying many different methods and optimization techniques. In contrast to the NSE results, we see that all methods and optimization techniques consistently outperform default RF when KGE is the metric of interest. Though all methods occasionally perform worse than default RF, all except for the optimal default RF method can outperform default RF by over 0.2 KGE on at least one dataset. This shows that default RF should never be used without checking other methods when KGE is the objective. The median increase in KGE for XGB with our meta-model is close to 0.1, while other median increases are closer to 0.05.

\subsection{New Optimal Defaults}

The new optimal defaults useful for Random Forest with KGE as an evaluation metric are as follows: mtry=0.885, numtrees=780, replace=True, min\_node\_size=0.14, and sample\_fraction=0.900. When using Random Forest with NSE as an evaluation metric the new optimal defaults we found were as follows: mtry=0.650, numtrees=49, replace=False, min\_nodesize=0.37, and  sample\_fraction=0.854. No matter the evaluation metric, our optimal default for mtry (0.885 or 0.65) is consistently higher than the optimal default from \citet{probst2019tunability}. While the original optimal default for the number of trees was found to be quite large at 983, we found that 780 trees can be optimal for KGE and 49 trees can be optimal for NSE. The previous optimal default recommends sampling without replacement (similar to the new NSE optimal default) while the default for the ranger package is to sample with replacement (similar to the new KGE optimal default). Both of our new optimal defaults for sample fraction (0.9 and 0.854) are greater than the old optimal default (0.703).

The new optimal defaults for XGBoost with KGE as an evaluation metric are as follows: numrounds=2945, eta=0.018, subsample=0.483, max\_depth=5, min\_child\_weight=1.380, colsample\_bytree=0.865, lambda=48.041, and alpha=0.019. The new optimal defaults for XGBoost with an NSE evaluation metric are as follows: numrounds=4942, eta=0.029, subsample=0.765, max\_depth=6, min\_child\_weight=6.156, colsample\_bytree=0.828, lambda=1019.458, and alpha=0.080. When comparing with the original optimal default, we see that numrounds is approximately the same when using NSE, while being significantly lower for KGE, however, eta was seen to be closer to the original using KGE. With NSE as an evaulation metric, the optimal subsample parameter (0.765) is only slightly less than the original optimal default (0.839), but KGE requires half (0.483). The optimal max\_depth was quite different in both cases (5 or 6 versus 13), but the NSE one was slightly closer. The min\_child\_weight value for the old optimal default was seen to be more similar with KGE, though considering the entire random search space, all three values are quite similar. Colsample\_bytree values are similar across all three optimal default estimates. Finally, the new optimal defaults for lambda (alpha) were significantly higher (lower), compared to the old optimal defaults.

\section{Conclusion}

One of the major challenges that water scientists face is model selection. This can be a major obstacle in the way of utilizing the full power of machine learning due to either the lack of knowledge or time. In this paper, we took a step towards removing these obstacles. We first compiled a database of 151 large-sample datasets coming from around the world. We then tested XGBoost and random forests paired with different hyperparameter optimization schemes and hydrologically relevant evaluation metrics (KGE or NSE). Using these results along with meta features from each dataset, we trained a meta model used to predict how a set of hyperparameters would perform given a new or arbitrary dataset. We then compared the predictive performance and time complexity of these methods. We found that xgboost is consistently faster than randomForest, but slower than ranger. Further, we showed that the NSE and KGE results tend to oppose each other where XGB performs best when KGE is the evaluation metric, while RF is slightly better when NSE is the evaluation metric. By providing water scientists with clear results which point to how to consistently outperform default random forest regression, they now have access to more robust and accurate methods for their analyses.

Machine learning is an ever growing field and the hyperparameter optimization techniques used in this field will be continually improved upon. To build on the results of this paper in future work we should compare new and recent hyperparameter optimization methods. The datasets we use in our meta-model are fairly limited, so extrapolating our results outside the training range of dataset characteristics may be unwise. We can run these methods on even more datasets to expand on the generality of our results. For example, we could use the CABra dataset which is a newly published large-sample hydrology dataset from Brazil or the CCAM dataset from China \citep{hao2021ccam,almagro2021cabra}. Further, we are looking forward to the development and release of other large sample hydrology datasets such as CAMELS\_FR and CAMELS\-spat \citep{knoben2022camels,andreassian2021camels}.

\section*{Acknowledgments}
Author contributions: J.J. supervised and designed the research; J.J., D.B. A.G., and A.M. performed the research; J.J., A.G., D.B., and A.M. wrote the paper; A.M., and J.J. did model coding, and D.B. and A.G. gave programming support. All code is available at \url{https://github.com/HydroML/HydroHyperOpt}.

\bibliographystyle{apalike}  
\bibliography{references}

\begin{thebibliography}{}

\bibitem[Addor et~al., 2020]{addor2020large}
Addor, N., Do, H.~X., Alvarez-Garreton, C., Coxon, G., Fowler, K., and Mendoza,
  P.~A. (2020).
\newblock Large-sample hydrology: recent progress, guidelines for new datasets
  and grand challenges.
\newblock {\em Hydrological Sciences Journal}, 65(5):712--725.

\bibitem[Addor et~al., 2018]{addor2018ranking}
Addor, N., Nearing, G., Prieto, C., Newman, A., Le~Vine, N., and Clark, M.~P.
  (2018).
\newblock A ranking of hydrological signatures based on their predictability in
  space.
\newblock {\em Water Resources Research}, 54(11):8792--8812.

\bibitem[Addor et~al., 2017]{addor2017camels}
Addor, N., Newman, A.~J., Mizukami, N., and Clark, M.~P. (2017).
\newblock The camels data set: catchment attributes and meteorology for
  large-sample studies.
\newblock {\em Hydrology and Earth System Sciences}, 21(10):5293--5313.

\bibitem[Alcoba{\c{c}}a et~al., 2020]{alcobacca2020mfe}
Alcoba{\c{c}}a, E., Siqueira, F., Rivolli, A., Garcia, L. P.~F., Oliva, J.~T.,
  de~Carvalho, A.~C., et~al. (2020).
\newblock Mfe: Towards reproducible meta-feature extraction.
\newblock {\em J. Mach. Learn. Res.}, 21(111):1--5.

\bibitem[Almagro et~al., 2021]{almagro2021cabra}
Almagro, A., Oliveira, P. T.~S., Meira~Neto, A.~A., Roy, T., and Troch, P.
  (2021).
\newblock Cabra: a novel large-sample dataset for brazilian catchments.
\newblock {\em Hydrology and Earth System Sciences}, 25(6):3105--3135.

\bibitem[Alvarez-Garreton et~al., 2018]{alvarez2018camels}
Alvarez-Garreton, C., Mendoza, P.~A., Boisier, J.~P., Addor, N., Galleguillos,
  M., Zambrano-Bigiarini, M., Lara, A., Puelma, C., Cortes, G., Garreaud, R.,
  et~al. (2018).
\newblock The camels-cl dataset: catchment attributes and meteorology for large
  sample studies--chile dataset.
\newblock {\em Hydrology and Earth System Sciences}, 22(11):5817--5846.

\bibitem[Andr{\'e}assian et~al., 2021]{andreassian2021camels}
Andr{\'e}assian, V., Delaigue, O., Perrin, C., Janet, B., and Addor, N. (2021).
\newblock Camels-fr: A large sample, hydroclimatic dataset for france, to
  support model testing and evaluation.
\newblock In {\em EGU General Assembly Conference Abstracts}, pages
  EGU21--13349.

\bibitem[Andr{\'e}assian et~al., 2010]{andreassian2010court}
Andr{\'e}assian, V., Perrin, C., Parent, E., and B{\'a}rdossy, A. (2010).
\newblock The court of miracles of hydrology: can failure stories contribute to
  hydrological science?

\bibitem[Ba{\c{s}}a{\u{g}}ao{\u{g}}lu et~al.,
  2020]{bacsaugaouglu2020hybridized}
Ba{\c{s}}a{\u{g}}ao{\u{g}}lu, H., Chakraborty, D., and Winterle, J. (2020).
\newblock A hybridized ngboost-xgboost framework for robust evaporation and
  evapotranspiration prediction.
\newblock {\em Hydrology and Earth System Sciences Discussions}, pages 1--27.

\bibitem[Bergstra and Bengio, 2012]{bergstra2012random}
Bergstra, J. and Bengio, Y. (2012).
\newblock Random search for hyper-parameter optimization.
\newblock {\em Journal of machine learning research}, 13(2).

\bibitem[Beven, 1989]{beven1989changing}
Beven, K. (1989).
\newblock Changing ideas in hydrology—the case of physically-based models.
\newblock {\em Journal of hydrology}, 105(1-2):157--172.

\bibitem[Bl{\"o}schl et~al., 2013]{bloschl2013runoff}
Bl{\"o}schl, G., Bloschl, G., Sivapalan, M., Wagener, T., Savenije, H., and
  Viglione, A. (2013).
\newblock {\em Runoff prediction in ungauged basins: synthesis across
  processes, places and scales}.
\newblock Cambridge University Press.

\bibitem[Breiman, 2001]{breiman2001random}
Breiman, L. (2001).
\newblock Random forests.
\newblock {\em Machine learning}, 45(1):5--32.

\bibitem[Browne, 2000]{browne2000cross}
Browne, M.~W. (2000).
\newblock Cross-validation methods.
\newblock {\em Journal of mathematical psychology}, 44(1):108--132.

\bibitem[Catav et~al., 2021]{catav2021marginal}
Catav, A., Fu, B., Zoabi, Y., Meilik, A. L.~W., Shomron, N., Ernst, J.,
  Sankararaman, S., and Gilad-Bachrach, R. (2021).
\newblock Marginal contribution feature importance-an axiomatic approach for
  explaining data.
\newblock In {\em International Conference on Machine Learning}, pages
  1324--1335. PMLR.

\bibitem[Chagas et~al., 2020]{chagas2020camels}
Chagas, V.~B., Chaffe, P.~L., Addor, N., Fan, F.~M., Fleischmann, A.~S., Paiva,
  R.~C., and Siqueira, V.~A. (2020).
\newblock Camels-br: hydrometeorological time series and landscape attributes
  for 897 catchments in brazil.
\newblock {\em Earth System Science Data}, 12(3):2075--2096.

\bibitem[Chang et~al., 2017]{chang2017determinants}
Chang, H., Bonnette, M.~R., Stoker, P., Crow-Miller, B., and Wentz, E. (2017).
\newblock Determinants of single family residential water use across scales in
  four western us cities.
\newblock {\em Science of the Total Environment}, 596:451--464.

\bibitem[Chen and Guestrin, 2016]{chen2016xgboost}
Chen, T. and Guestrin, C. (2016).
\newblock Xgboost: A scalable tree boosting system.
\newblock In {\em Proceedings of the 22nd acm sigkdd international conference
  on knowledge discovery and data mining}, pages 785--794.

\bibitem[Chen et~al., 2015]{chen2015xgboost}
Chen, T., He, T., Benesty, M., Khotilovich, V., Tang, Y., Cho, H., Chen, K.,
  et~al. (2015).
\newblock Xgboost: extreme gradient boosting.
\newblock {\em R package version 0.4-2}, 1(4):1--4.

\bibitem[Covert et~al., 2020]{covert2020understanding}
Covert, I., Lundberg, S.~M., and Lee, S.-I. (2020).
\newblock Understanding global feature contributions with additive importance
  measures.
\newblock {\em Advances in Neural Information Processing Systems},
  33:17212--17223.

\bibitem[Coxon et~al., 2020]{coxon2020camels}
Coxon, G., Addor, N., Bloomfield, J.~P., Freer, J., Fry, M., Hannaford, J.,
  Howden, N.~J., Lane, R., Lewis, M., Robinson, E.~L., et~al. (2020).
\newblock Camels-gb: Hydrometeorological time series and landscape attributes
  for 671 catchments in great britain.
\newblock {\em Earth System Science Data}, 12(4):2459--2483.

\bibitem[Cui et~al., 2016]{cui2016recommendation}
Cui, C., Hu, M., Weir, J.~D., and Wu, T. (2016).
\newblock A recommendation system for meta-modeling: A meta-learning based
  approach.
\newblock {\em Expert Systems with Applications}, 46:33--44.

\bibitem[Fan et~al., 2018]{fan2018evaluation}
Fan, J., Yue, W., Wu, L., Zhang, F., Cai, H., Wang, X., Lu, X., and Xiang, Y.
  (2018).
\newblock Evaluation of svm, elm and four tree-based ensemble models for
  predicting daily reference evapotranspiration using limited meteorological
  data in different climates of china.
\newblock {\em Agricultural and forest meteorology}, 263:225--241.

\bibitem[Feigl et~al., 2021]{feigl2021machine}
Feigl, M., Lebiedzinski, K., Herrnegger, M., and Schulz, K. (2021).
\newblock Machine-learning methods for stream water temperature prediction.
\newblock {\em Hydrology and Earth System Sciences}, 25(5):2951--2977.

\bibitem[Feurer and Hutter, 2019]{feurer2019hyperparameter}
Feurer, M. and Hutter, F. (2019).
\newblock Hyperparameter optimization.
\newblock In {\em Automated machine learning}, pages 3--33. Springer, Cham.

\bibitem[Feurer et~al., 2015]{feurer2015efficient}
Feurer, M., Klein, A., Eggensperger, K., Springenberg, J., Blum, M., and
  Hutter, F. (2015).
\newblock Efficient and robust automated machine learning.
\newblock {\em Advances in neural information processing systems}, 28.

\bibitem[Fowler et~al., 2021]{fowler2021camels}
Fowler, K.~J., Acharya, S.~C., Addor, N., Chou, C., and Peel, M.~C. (2021).
\newblock Camels-aus: hydrometeorological time series and landscape attributes
  for 222 catchments in australia.
\newblock {\em Earth System Science Data}, 13(8):3847--3867.

\bibitem[Gauch et~al., 2019]{gauch2019data}
Gauch, M., Mai, J., Gharari, S., and Lin, J. (2019).
\newblock Data-driven vs. physically-based streamflow prediction models.
\newblock In {\em Proceedings of 9th International Workshop on Climate
  Informatics}.

\bibitem[Gauch et~al., 2021]{gauch2021proper}
Gauch, M., Mai, J., and Lin, J. (2021).
\newblock The proper care and feeding of camels: How limited training data
  affects streamflow prediction.
\newblock {\em Environmental Modelling \& Software}, 135:104926.

\bibitem[Golden et~al., 2017]{golden2017integrating}
Golden, H.~E., Creed, I.~F., Ali, G., Basu, N.~B., Neff, B.~P., Rains, M.~C.,
  McLaughlin, D.~L., Alexander, L.~C., Ameli, A.~A., Christensen, J.~R., et~al.
  (2017).
\newblock Integrating geographically isolated wetlands into land management
  decisions.
\newblock {\em Frontiers in Ecology and the Environment}, 15(6):319--327.

\bibitem[Gupta et~al., 2009]{gupta2009decomposition}
Gupta, H.~V., Kling, H., Yilmaz, K.~K., and Martinez, G.~F. (2009).
\newblock Decomposition of the mean squared error and nse performance criteria:
  Implications for improving hydrological modelling.
\newblock {\em Journal of hydrology}, 377(1-2):80--91.

\bibitem[Gupta et~al., 2014]{gupta2014large}
Gupta, H.~V., Perrin, C., Bl{\"o}schl, G., Montanari, A., Kumar, R., Clark, M.,
  and Andr{\'e}assian, V. (2014).
\newblock Large-sample hydrology: a need to balance depth with breadth.
\newblock {\em Hydrology and Earth System Sciences}, 18(2):463--477.

\bibitem[Halford, 2004]{halford2004more}
Halford, K.~J. (2004).
\newblock More data required.
\newblock {\em Ground Water}, 42(4):477--478.

\bibitem[Hao et~al., 2021]{hao2021ccam}
Hao, Z., Jin, J., Xia, R., Tian, S., Yang, W., Liu, Q., Zhu, M., Ma, T., Jing,
  C., and Zhang, Y. (2021).
\newblock Ccam: China catchment attributes and meteorology dataset.
\newblock {\em Earth System Science Data}, 13(12):5591--5616.

\bibitem[Hastie et~al., 2009]{hastie2009random}
Hastie, T., Tibshirani, R., and Friedman, J. (2009).
\newblock Random forests.
\newblock In {\em The elements of statistical learning}, pages 587--604.
  Springer.

\bibitem[Hrachowitz et~al., 2013]{hrachowitz2013decade}
Hrachowitz, M., Savenije, H., Bl{\"o}schl, G., McDonnell, J., Sivapalan, M.,
  Pomeroy, J., Arheimer, B., Blume, T., Clark, M., Ehret, U., et~al. (2013).
\newblock A decade of predictions in ungauged basins (pub)—a review.
\newblock {\em Hydrological sciences journal}, 58(6):1198--1255.

\bibitem[Janssen and Ameli, 2021]{janssen2021hydrologic}
Janssen, J. and Ameli, A.~A. (2021).
\newblock A hydrologic functional approach for improving large-sample hydrology
  performance in poorly gauged regions.
\newblock {\em Water Resources Research}, 57(9):e2021WR030263.

\bibitem[Janssen et~al., 2023]{janssen2022ultra}
Janssen, J., Guan, V., and Robeva, E. (2023).
\newblock Ultra-marginal feature importance: Learning from data with causal
  guarantees.
\newblock In {\em International Conference on Artificial Intelligence and
  Statistics}, pages 10782--10814. PMLR.

\bibitem[Kling et~al., 2012]{kling2012runoff}
Kling, H., Fuchs, M., and Paulin, M. (2012).
\newblock Runoff conditions in the upper danube basin under an ensemble of
  climate change scenarios.
\newblock {\em Journal of Hydrology}, 424:264--277.

\bibitem[Klingler et~al., 2021]{klingler2021lamah}
Klingler, C., Schulz, K., and Herrnegger, M. (2021).
\newblock Lamah-ce: Large-sample data for hydrology and environmental sciences
  for central europe.
\newblock {\em Earth System Science Data}, 13(9):4529--4565.

\bibitem[Knoben and Clark, 2022]{knoben2022camels}
Knoben, W. and Clark, M. (2022).
\newblock Camels-spat: catchment data for spatially distributed large-sample
  hydrology.
\newblock Technical report, Copernicus Meetings.

\bibitem[Kuentz et~al., 2017]{kuentz2017understanding}
Kuentz, A., Arheimer, B., Hundecha, Y., and Wagener, T. (2017).
\newblock Understanding hydrologic variability across europe through catchment
  classification.
\newblock {\em Hydrology and Earth System Sciences}, 21(6):2863--2879.

\bibitem[Lange and Sippel, 2020]{lange2020machine}
Lange, H. and Sippel, S. (2020).
\newblock Machine learning applications in hydrology.
\newblock In {\em Forest-water interactions}, pages 233--257. Springer.

\bibitem[LeDell and Poirier, 2020]{ledell2020h2o}
LeDell, E. and Poirier, S. (2020).
\newblock H2o automl: Scalable automatic machine learning.
\newblock In {\em Proceedings of the AutoML Workshop at ICML}, volume 2020.

\bibitem[Li and Ameli, 2022]{li2022statistical}
Li, H. and Ameli, A. (2022).
\newblock A statistical approach for identifying factors governing streamflow
  recession behaviour.
\newblock {\em Hydrological Processes}, 36(10):e14718.

\bibitem[Liaw et~al., 2002]{liaw2002classification}
Liaw, A., Wiener, M., et~al. (2002).
\newblock Classification and regression by randomforest.
\newblock {\em R news}, 2(3):18--22.

\bibitem[Linsley, 1982]{linsley1982rainfall}
Linsley, R. (1982).
\newblock Rainfall-runoff models-an overview.
\newblock In {\em Proceedings of the international symposium on rainfall-runoff
  modelling, edited by: Singh, VP, Water Resources Publications, Littleton,
  CO}, pages 3--22.

\bibitem[Lorena et~al., 2019]{lorena2019complex}
Lorena, A.~C., Garcia, L.~P., Lehmann, J., Souto, M.~C., and Ho, T.~K. (2019).
\newblock How complex is your classification problem? a survey on measuring
  classification complexity.
\newblock {\em ACM Computing Surveys (CSUR)}, 52(5):1--34.

\bibitem[Majumdar et~al., 2020]{majumdar2020groundwater}
Majumdar, S., Smith, R., Butler~Jr, J., and Lakshmi, V. (2020).
\newblock Groundwater withdrawal prediction using integrated multitemporal
  remote sensing data sets and machine learning.
\newblock {\em Water Resources Research}, 56(11):e2020WR028059.

\bibitem[McCuen et~al., 2006]{mccuen2006evaluation}
McCuen, R.~H., Knight, Z., and Cutter, A.~G. (2006).
\newblock Evaluation of the nash--sutcliffe efficiency index.
\newblock {\em Journal of hydrologic engineering}, 11(6):597--602.

\bibitem[Nash and Sutcliffe, 1970]{nash1970river}
Nash, J.~E. and Sutcliffe, J.~V. (1970).
\newblock River flow forecasting through conceptual models part i—a
  discussion of principles.
\newblock {\em Journal of hydrology}, 10(3):282--290.

\bibitem[Ni et~al., 2020]{ni2020streamflow}
Ni, L., Wang, D., Wu, J., Wang, Y., Tao, Y., Zhang, J., and Liu, J. (2020).
\newblock Streamflow forecasting using extreme gradient boosting model coupled
  with gaussian mixture model.
\newblock {\em Journal of Hydrology}, 586:124901.

\bibitem[Ntokas et~al., 2021]{ntokas2021investigating}
Ntokas, K.~F., Odry, J., Boucher, M.-A., and Garnaud, C. (2021).
\newblock Investigating ann architectures and training to estimate snow water
  equivalent from snow depth.
\newblock {\em Hydrology and Earth System Sciences}, 25(6):3017--3040.

\bibitem[Papacharalampous et~al., 2018]{papacharalampous2018univariate}
Papacharalampous, G., Tyralis, H., and Koutsoyiannis, D. (2018).
\newblock Univariate time series forecasting of temperature and precipitation
  with a focus on machine learning algorithms: A multiple-case study from
  greece.
\newblock {\em Water resources management}, 32(15):5207--5239.

\bibitem[Pham et~al., 2021]{pham2021evaluation}
Pham, L.~T., Luo, L., and Finley, A. (2021).
\newblock Evaluation of random forests for short-term daily streamflow
  forecasting in rainfall-and snowmelt-driven watersheds.
\newblock {\em Hydrology and Earth System Sciences}, 25(6):2997--3015.

\bibitem[Probst et~al., 2019]{probst2019tunability}
Probst, P., Boulesteix, A.-L., and Bischl, B. (2019).
\newblock Tunability: importance of hyperparameters of machine learning
  algorithms.
\newblock {\em The Journal of Machine Learning Research}, 20(1):1934--1965.

\bibitem[Saia et~al., 2020]{saia2020transitioning}
Saia, S.~M., Nelson, N., Huseth, A.~S., Grieger, K., and Reich, B.~J. (2020).
\newblock Transitioning machine learning from theory to practice in natural
  resources management.
\newblock {\em Ecological Modelling}, 435:109257.

\bibitem[Sekuli{\'c} et~al., 2021]{sekulic2021high}
Sekuli{\'c}, A., Kilibarda, M., Proti{\'c}, D., and Bajat, B. (2021).
\newblock A high-resolution daily gridded meteorological dataset for serbia
  made by random forest spatial interpolation.
\newblock {\em Scientific Data}, 8(1):1--12.

\bibitem[Svetnik et~al., 2003]{svetnik2003random}
Svetnik, V., Liaw, A., Tong, C., Culberson, J.~C., Sheridan, R.~P., and
  Feuston, B.~P. (2003).
\newblock Random forest: a classification and regression tool for compound
  classification and qsar modeling.
\newblock {\em Journal of chemical information and computer sciences},
  43(6):1947--1958.

\bibitem[Tang et~al., 2020]{tang2020scdna}
Tang, G., Clark, M.~P., Newman, A.~J., Wood, A.~W., Papalexiou, S.~M., Vionnet,
  V., and Whitfield, P.~H. (2020).
\newblock Scdna: A serially complete precipitation and temperature dataset for
  north america from 1979 to 2018.
\newblock {\em Earth System Science Data}, 12(4):2381--2409.

\bibitem[Teweldebrhan et~al., 2020]{teweldebrhan2020coupled}
Teweldebrhan, A.~T., Schuler, T.~V., Burkhart, J.~F., and Hjorth-Jensen, M.
  (2020).
\newblock Coupled machine learning and the limits of acceptability approach
  applied in parameter identification for a distributed hydrological model.
\newblock {\em Hydrology and Earth System Sciences}, 24(9):4641--4658.

\bibitem[Thornton et~al., 2013]{thornton2013auto}
Thornton, C., Hutter, F., Hoos, H.~H., and Leyton-Brown, K. (2013).
\newblock Auto-weka: Combined selection and hyperparameter optimization of
  classification algorithms.
\newblock In {\em Proceedings of the 19th ACM SIGKDD international conference
  on Knowledge discovery and data mining}, pages 847--855.

\bibitem[Tyralis et~al., 2019]{tyralis2019brief}
Tyralis, H., Papacharalampous, G., and Langousis, A. (2019).
\newblock A brief review of random forests for water scientists and
  practitioners and their recent history in water resources.
\newblock {\em Water}, 11(5):910.

\bibitem[Tyralis et~al., 2021]{tyralis2021super}
Tyralis, H., Papacharalampous, G., and Langousis, A. (2021).
\newblock Super ensemble learning for daily streamflow forecasting: Large-scale
  demonstration and comparison with multiple machine learning algorithms.
\newblock {\em Neural Computing and Applications}, 33(8):3053--3068.

\bibitem[Vanschoren, 2019]{vanschoren2019meta}
Vanschoren, J. (2019).
\newblock Meta-learning.
\newblock In {\em Automated Machine Learning}, pages 35--61. Springer, Cham.

\bibitem[Worland et~al., 2018]{worland2018improving}
Worland, S.~C., Farmer, W.~H., and Kiang, J.~E. (2018).
\newblock Improving predictions of hydrological low-flow indices in ungaged
  basins using machine learning.
\newblock {\em Environmental modelling \& software}, 101:169--182.

\bibitem[Wright and Ziegler, 2015]{wright2015ranger}
Wright, M.~N. and Ziegler, A. (2015).
\newblock ranger: A fast implementation of random forests for high dimensional
  data in c++ and r.
\newblock {\em arXiv preprint arXiv:1508.04409}.

\bibitem[Yang and Shami, 2020]{yang2020hyperparameter}
Yang, L. and Shami, A. (2020).
\newblock On hyperparameter optimization of machine learning algorithms: Theory
  and practice.
\newblock {\em Neurocomputing}, 415:295--316.

\bibitem[Zhang et~al., 2019]{zhang2019dynamic}
Zhang, H., Yang, Q., Shao, J., and Wang, G. (2019).
\newblock Dynamic streamflow simulation via online gradient-boosted regression
  tree.
\newblock {\em Journal of Hydrologic Engineering}, 24(10):04019041.

\bibitem[Zhang et~al., 2021]{zhang2021global}
Zhang, Y., Wang, X., Pan, Y., Hu, R., and Chen, N. (2021).
\newblock Global quantitative synthesis of effects of biotic and abiotic
  factors on stemflow production in woody ecosystems.

\bibitem[Z{\"o}ller and Huber, 2021]{zoller2021benchmark}
Z{\"o}ller, M.-A. and Huber, M.~F. (2021).
\newblock Benchmark and survey of automated machine learning frameworks.
\newblock {\em Journal of artificial intelligence research}, 70:409--472.

\end{thebibliography}

\end{document}